\documentclass[aps,prapplied,twocolumn,groupedaddress,floatfix,longbibliography]{revtex4-2}

\usepackage{graphicx}
\usepackage{dcolumn}
\usepackage{bm}
\usepackage{amsmath}
\usepackage{xcolor}
\usepackage{braket}
\usepackage{sidecap}
\usepackage{wrapfig}
\usepackage[utf8]{inputenc}
\usepackage[T1]{fontenc}
\usepackage{hyperref}
\hypersetup{
breaklinks=true,
    colorlinks=true,
    linkcolor=blue,
    citecolor=blue,
    filecolor=magenta,
    urlcolor=cyan,
}

\usepackage{blkarray}
\usepackage{multirow}

\usepackage{natbib}      

\renewcommand{\ae}{{\alpha\eta}}
\newcommand{\az}{{\alpha\zeta}}
\newcommand{\bz}{{\beta\zeta}}
\newcommand{\be}{{\beta\eta}}

\newcommand{\la}{\lambda}
\newcommand{\si}{\sigma}

\newcommand{\ve}{\varepsilon}
\newcommand{\vp}{\varphi}

\newcommand{\uhf}{ \Psi_{\rm UHF} }
\newcommand{\puhf}{ \Psi_{\rm PUHF} }
\newcommand{\cpy}{\hat{\cal P}_y}
\newcommand{\hpi}{\hat{\Pi}}

\newcommand{\br}{{\bf r}}

\begin{document}

\title{Electronic Wigner-Molecule Polymeric Chains in Elongated Silicon Quantum Dots and Finite-Length Quantum Wires} 

\author{Arnon Goldberg}
\email{agoldberg36@gatech.edu}
\author{Constantine Yannouleas}
\email{Constantine.Yannouleas@physics.gatech.edu}
\author{Uzi Landman}
\email{Uzi.Landman@physics.gatech.edu}

\affiliation{School of Physics, Georgia Institute of Technology,
             Atlanta, Georgia 30332-0430}

\date{03 February 2024}

\begin{abstract}
The spectral properties of electrons confined in a wire-like quasi-one-dimensional (1D) elongated quantum 
dot (EQD) coupler between silicon qubits, are investigated with a newly developed valley-augmented  
unrestricted Hartree-Fock (va-UHF) method, generalized to include the valley degree of freedom 
\textcolor{black}{treated as an isospin}, allowing 
calculations for a large number of electrons. The lower energy symmetry-broken solutions of the 
self-consistent generalized Pople-Nesbet equations exhibit, for a confinement that has been modeled after 
an experimentally fabricated one in silicon, formation of Wigner-molecular polymeric (longitudinal)
chains, initiating through charge accumulation at the 
edges of the finite-length quasi-1D wire. An increasing number of 
parallel zig-zag chains form as the number of electrons loaded into the confinement is increased, 
with the formation of newly added chains determined by the strength of the transverse harmonic
confinement. The broken-symmetry va-UHF solutions, subsequently augmented by the quantum-mechanically 
required parity-restoration, go beyond the va-UHF single-determinant solution, predicting formation of 
entangled Wigner-molecular chains whose charge distributions obliterate the zig-zag organization of the  
broken-symmetry solutions. The symmetry-restored va-UHF methodology enables systematic investigations of 
multi-electron complex nano-scale confined structures that could be targeted for future imaging 
microscopy experiments in silicon and other  materials (e.g., 1D domain walls in TMD materials),   
and quantum information utilization. 
\end{abstract}   

\maketitle

\section{Introduction} 

Quantum-dot (QD) qubits are fundamental elements for semiconductor-based 
solid-state quantum computing architectures \cite{dzur13,warr22,burk23}. 
One of the central challenging issues in constructing scalable quantum processors is that of 
quantum chip large-scale integration, allowing transfer of information between computing qubits 
while preserving information during transfer. 

Currently, attention focuses on patterned, gate-controlled elongated quantum dots (EQDs) enabling coherent
transfer of spins between relatively distant quantum-dot qubits \cite{kuem17,kuem18,dzur23,kuem23}, 
thus reducing the complexity and technical difficulties that accompany short distance multi-dot couplers. 
To overcome the challenges of designing qubit coupling strategies, recent developments \cite{dzur23,kuem23}
focus on silicon-based nanodevices \cite{dzur13,warr22} guided by: (i) the 
long coherence gained via the use of enriched $^{28}$Si substrates, (ii) the successful 
demonstrations of Si-based high-speed operation and high fidelity spin quantum qubits, and (iii) the
vast investment made in the industry and the ensuing infrastructure availability, as well as the
scientific experience gained already with Si-based technologies.
 
Here we aim at gaining fundamental insights about the 
many-body quantum nature of the electronic states in such patterned, long-distance-coupling
EQDs. Such understanding is imperative for enabling theory-guided fabrication and 
integration of these elements into solid-state Si-based quantum information devices. Our main
finding is that the extended nature of the (wire-like) EQD results in conditions where the
inter-electron repulsion energy dominates over the electron quantal kinetic energy. 
\textcolor{black}{These conditions underlie the emergent formation} 
of pinned Wigner Molecules (WMs), exhibiting a general architecture of parallel chains. 
Quantum WMs (manifesting sliding \footnote{In the case of a confinement with circular symmetry, the 
WMs are referred to as rotating WMs as well} 
or pinned geometrical configurations) have been predicted theoretically (see, e.g., 
Refs.\ \cite{yann99,yann00,yann02.2,harj02,mikh02,yann03,szaf03,yann07,yann11}), and subsequently 
confirmed experimentally in several materials systems (Ga[Al]As single QDs \cite{yann06,ihn07},
GaAs double QDs \cite{kim21,yann22,yann22.2}, 
Si QDs \cite{corr21,yann22.3}, carbon nanotubes \cite{peck13}), and most recently, 
in agreement with the latest theoretical predictions \cite{yann23,yann24}, in moir\'e superstructures in 
transition metal dichalcogenide (TMD) materials \cite{crom23}.

Naturally, one proceeds by formulating the many-body Hamiltonian describing the wire-like 
EQD, and by numerically solving the corresponding Schr\"{o}dinger equation. 
The most accurate method to this effect is exact diagonalization (EXD) through the use of full 
configuration interaction (FCI) \cite{lowd55,shav98,yann03,yann07,yann22.2,yann22.3}, which 
is limited to systems containing up to about  10 confined electrons,
whereas the systems of relevance for qubit couplers are expected to consist of a much larger number 
of electrons. Accordingly, we employ the unrestricted Hartree-Fock (UHF) 
\cite{szabo,yann99,yann02.2,yann07}, 
which refers to a family of self-consistent field calculation techniques. 

The UHF methodology \textcolor{black}{uncovers formation of}
crystalline-like space-symmetry-broken charge densities (CDs), 
which are finite-size analogs of the Wigner-crystal chains investigated previously theoretically 
\textcolor{black}{in the classical limit \cite{piac04} and in the context of Heisenberg chain modeling 
\cite{matv08}, as well as investigated}
experimentally for very long GaAs/AlGaAs wires \cite{pepp18,pepp21}. 

A proper quantum mechanical description requires the restoration 
\cite{yann02.2,yann07,yann21} of the UHF broken symmetries with respect to all the symmetry 
operations commuting with the system Hamiltonian. Such symmetry restoration is implemented here with 
respect to the $y$-parity symmetry about the long $x$-axis of the EQD. We discovered that the 
parity-symmetry-restored solutions yield a substantially improved description of the CDs compared to 
those given by the broken-symmetry UHF solutions. Moreover, this improvement, involving an extension 
beyond the variational UHF single-determinant ansatz, proves to be most efficient, entailing a rather 
minimal computational cost for systems involving a relatively large number (dozens) of electrons.

Furthermore, the present implementation of the UHF method requires consideration of the valley degree 
of freedom which is present in silicon nanostructures, treated here as a pseudospin (along the spin).
This requires modifications (employed in this paper; see Sec.\ \ref{vauhf}) 
of the UHF Pople-Nesbet equations \cite{yann99,yann02.2,yann07} formulated originally in quantum chemistry 
for the case of natural molecules \cite{szabo}. We will interchangeably refer to our calculations 
as UHF or valley-augmented UHF (va-UHF).

\begin{figure}[b]
\centering\includegraphics[width=8.0cm]{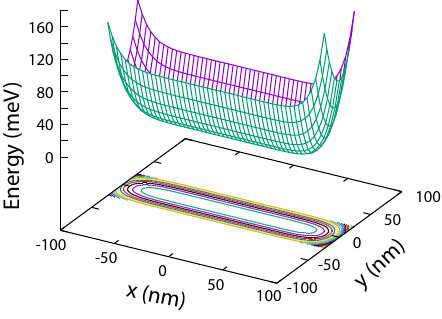}
\caption{
An illustration of the longboat-Viking-type total confining potential $U(x,y)$ used. 
The square-type lateral potential, $U_L(x)$, along the $x$-axis is given by Eq.\ (\ref{potx}). 
The transverse harmonic potential, $U_T(y)$, along 
the $y$-direction corresponds to $\hbar \omega_y=10$ meV and to an effective electron mass 
$m^*=0.19 m_e$ (appropriate for silicon).
}
\label{s1}
\end{figure}

\begin{figure}[b]
\centering\includegraphics[width=7.5cm]{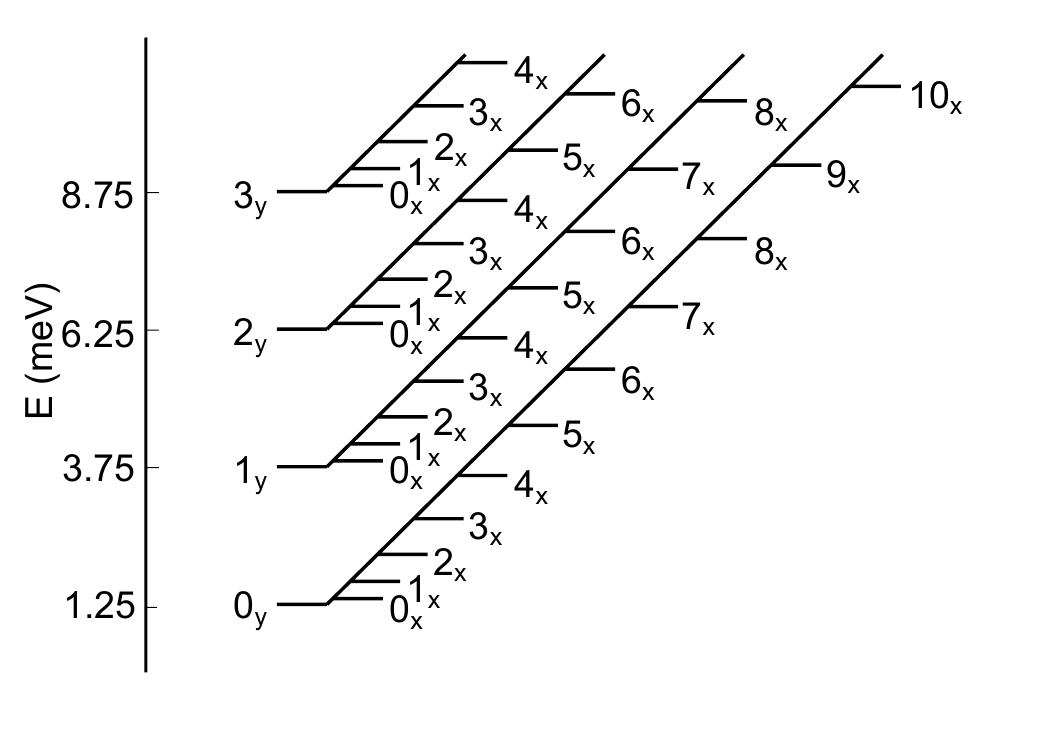}
\caption{
An illustration of the single-particle spectrum of electrons in the potential, $U(x,y)$, used to 
model the elongated QD investigated in this paper. For the parameters, see the text.
The labels for the single-particle space orbitals are as follows: $n_x \rightarrow$ number of nodes 
in the $x$-direction. $n_y \rightarrow$ number of nodes in the $y$-direction.}
\label{s2}
\end{figure}

\section{Many-body Hamiltonian}

We consider a square-like, slightly asymmetric, confining potential along the 
lateral long axis (the $x$-axis) of the EQD specified by: 
\begin{align}
U_L(x)=\xi |x/x_0|^9 \Theta(-x)+ \xi (x/x_0)^{10} \Theta(x),
\label{potx}
\end{align} 
where $\Theta(x)$ is the Heaviside step function, $x_0=50$ nm, and $\xi=0.1$ meV. This confinement 
corresponds to the EQD (referred to also as jellybean QD) fabricated (including the above-noted
slight asymmetry) and investigated in Ref.\ \cite{dzur23}. 
The total confinement is given by 
\begin{align}
U(x,y)=U_L(x)+U_T(y),
\label{poxy}
\end{align}
where $U_T(y)=m^*\omega_y^2 y^2/2$ is a harmonic potential along the transverse $y$-direction; 
$m^*=0.19m_e$ is the effective electron mass for Si and $\hbar \omega_y$ the harmonic quantum in 
the transverse $y$-direction. 

An illustration of $U(x,y)$ is presented in Fig.\ \ref{s1}. Furthermore, an illustration of the 
associated (single-particle) spectrum (with $\hbar \omega_y=2.5$ meV and $m^*=0.19 m_e$) is plotted 
in Fig.\ \ref{s2}.

The relevant many-body Hamiltonian is given by:
\begin{align}
H_{\rm MB} = \sum_{i=1}^N \left( \frac{{\bf p}_i^2}{2 m^*} + U(\br_i) \right) +
\frac{e^2}{\kappa} \sum_{i=1}^{N-1} \sum_{j>i}^N \frac{1}{|\br_i-\br_j|},
\label{mbh}
\end{align} 
where $\kappa$ is the dielectric constant.

\begin{figure*}[t]
\centering\includegraphics[width=12.0cm]{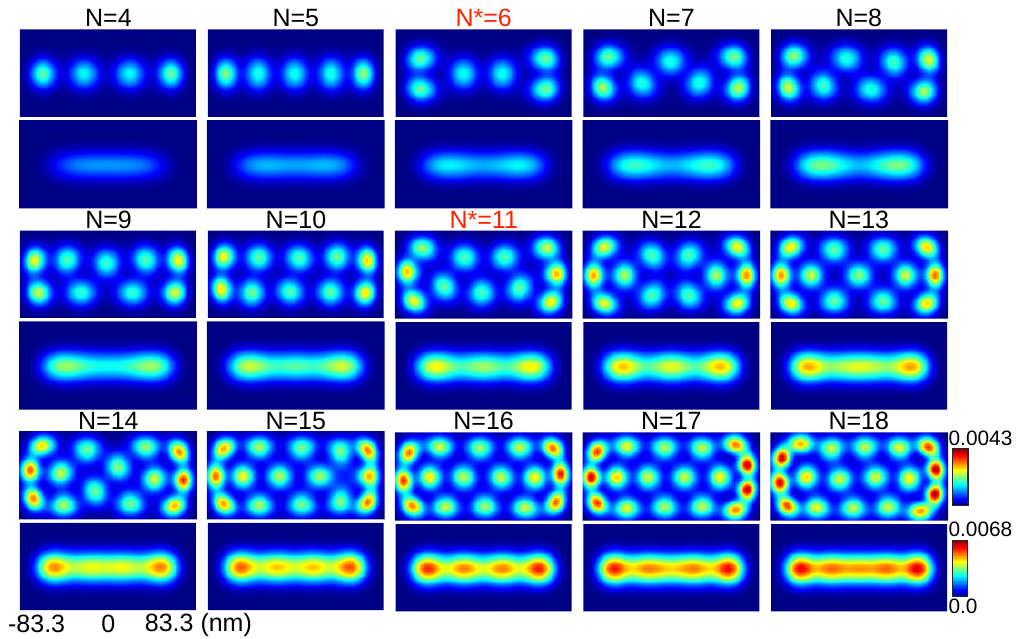}
\caption{
UHF ground-state charge densities (CDs) for $N=4-18$ electrons in a silicon elongated quantum dot 
($\sim 110$ nm long), with a transverse confinemnet of $\hbar \omega_y = 2.5$ meV. For each $N$ the CD 
is shown for strongly repelling  electrons ($\kappa =11$) at the top, and the CD for electrons in the 
non-interacting limit is shown at the bottom. The non-interacting CDs (at maximum occupation per orbital, 
large $\kappa=300$, and $\hbar \omega_y=2.5$ meV) exhibit for all $N$ single-row 
structures of a delocalized-particle nature, and obey a shell-filling Aufbau rule (following spin and 
isospin exclusion rules; see text). In contrast, the strongly interacting electrons CDs exhibit a 
transition from linear Wigner-molecular structures for $N \leq 5$ to a single zig-zag chain at 
$N^*=5$ and to a double zig-zag chain at $N^*=11$, that initiate via charge accumulation occurring at 
the ends of the EQD. The color bars (bottom right) indicate the charge density scale (in units of
1/nm$^2$). For details, see the text.  
}
\label{uhfdens1}
\end{figure*}

\begin{figure*}[t]
\centering\includegraphics[width=14.0cm]{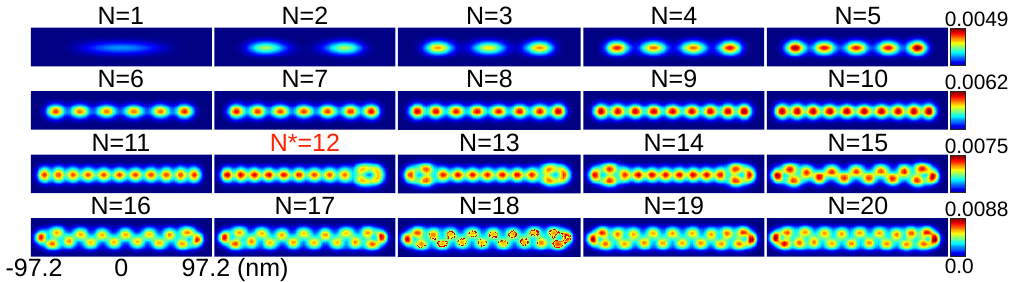}
\caption{
UHF ground-state CDs for $N=1-20$ electrons for a silicon EQD with a transverse 
confinement of $\hbar \omega_y=10$ meV. $\kappa=11$ appropriate for silicon (strong interaction).
The CDs exhibit a transition from one-row Wigner-molecular structures for $N \leq 11$ to a zig-zag 
chain for $N \geq N^* = 12$ that initiates via charge accumulation occurring at the two ends of the 
EQD (see $N=12-14$), with the zig-zag pattern fully developed for $N = 15 - 20$. 
For larger number of electrons a three-row zig-zag chain structure emerges (not shown). The color 
bars (on the right) indicate the charge-density scale (in units of 1/nm$^2$).
}
\label{uhfdens2}
\end{figure*}

\section{THE VALLEY-AUGMENTED UNRESTRICTED HARTREE-FOCK FOR NANOSYSTEMS}
\label{vauhf}

Due to strain in Si/SiGe quantum wells and (interfacial) quantum dots (in particular, in heterostructure
semiconductors)  and higher subband quantization energy in metal-oxide-semiconductor devices, 
the energies of the (four) in-plane Si valleys are raised, resulting in two remaining degenerate valleys. 
In this paper, as is the case for the general practice for
the Si QDs \cite{copp13}, we consider that the valley degree of freedom (VDOF) consists only of the 
low-energy two-fold band.

As elaborated in Ref.\ \cite{yann22.3}, to characterize and classify the VDOF of the two remaining valleys,
we consider in this paper an isospin designation that is constructed in analogy [including the SU(2)
algebra generated by the $i$-multiplied Pauli matrices] with that of the regular spin of the electrons
(replacing $\hat{\bf S}$ with $\hat{\bf V}$ when referring to the VDOF). Obviously, in the absence of 
two-body interactions, the occupation of the single-particle energy states \footnote{
These confinement-induced single-particle energy states are also referred to as ``orbitals'', see, e.g.,
the expressions ``atomic orbitals'', ``space orbitals'', and ``spin-orbitals'' in chemistry \cite{szabo}}
in the QDs would depend on an interplay between the confinement (including an applied magnetic field) and 
the spin and valley effects. For the case of two-valley degeneracy (considered here) this interplay results
in ``quadrupling'' of the spectrum (four degenerate states for each confinement state, each  corresponding 
to a different spin and valley).

Here, we introduce the va-UHF for solving the many-body problem specified by the Hamiltonian $H_{\rm MB}$ 
defined in Eq.\ (\ref{mbh}). Accordingly, one introduces a basis set 
of spatial single-particle orbitals $\varphi_\mu (x,y)$, $\mu=1,\ldots,K$, 
which are given by the $K$ lowest-energy solutions of the auxiliary single-particle Hamiltonian,
\begin{align}
H^{\rm core} = \sum_{i=1}^N \left( \frac{{\bf p}_i^2}{2 m^*} + U(\br_i) \right).
\label{spo}
\end{align}
For clarity and convenience, these solutions are sorted in ascending energy.

Subsequently, for the four families of va-UHF {\it spin-isospin-orbitals\/} \footnote{
This is an apparent generalization of the term spin-orbital used in chemistry and molecular physics}, 
$\psi_i^\az \alpha\zeta$, $\psi_i^\bz \beta\zeta$, $\psi_i^\ae \alpha\eta$, 
$\psi_i^\be \beta\eta$, where $\alpha (\beta)$ denote up (down) spins, $\zeta (\eta)$ denote up 
(down) isospins [i.e., electrons in the first (second) valley], one considers the expansions:
\begin{align}
\begin{split}
& \psi_i^\az=\sum_{\mu=1}^K C_{\mu i}^\az \vp_\mu,\\
& \psi_i^\bz=\sum_{\mu=1}^K C_{\mu i}^\bz \vp_\mu, \\
& \psi_i^\ae=\sum_{\mu=1}^K C_{\mu i}^\ae \vp_\mu,\\
& \psi_i^\be=\sum_{\mu=1}^K C_{\mu i}^\be \vp_\mu, \\
& i=1,2,\ldots,K.
\end{split}\
\end{align}

Then, following similar steps as in Ch. 3.8.2 of Ref.\ \cite{szabo}, we have derived a generalization of the 
Pople-Nesbet equations as follows:
\begin{align}
\begin{split}
 & \sum_\nu F_{\mu \nu}^\az C_{\nu j}^\az =  \ve_j^\az \sum_\nu
  S_{\mu \nu}C_{\nu j}^\az,\\
 & \sum_\nu F_{\mu \nu}^\bz C_{\nu j}^\bz =  \ve_j^\bz \sum_\nu
  S_{\mu \nu}C_{\nu j}^\bz,\\
 & \sum_\nu F_{\mu \nu}^\ae C_{\nu j}^\ae =  \ve_j^\ae \sum_\nu
  S_{\mu \nu}C_{\nu j}^\ae,\\
 & \sum_\nu F_{\mu \nu}^\be C_{\nu j}^\be =  \ve_j^\be \sum_\nu
  S_{\mu \nu}C_{\nu j}^\be,\\
 & j=1,2,\ldots,K.
\label{popnes}
\end{split}
\end{align}
with the Fock-operator matrices being given by,
 \begin{align}
 \begin{split}
 F_{\mu\nu}^\az= H_{\mu\nu}^{\rm core}  +
\sum_{\la} \sum_{\si} P_{\la\si}^T [(\mu\nu|\si\la)- P_{\la\si}^\az (\mu\la|\si\nu),\\
 F_{\mu\nu}^\bz= H_{\mu\nu}^{\rm core}  +
\sum_{\la} \sum_{\si} P_{\la\si}^T [(\mu\nu|\si\la)- P_{\la\si}^\bz (\mu\la|\si\nu),\\
 F_{\mu\nu}^\ae= H_{\mu\nu}^{\rm core}  +
\sum_{\la} \sum_{\si} P_{\la\si}^T [(\mu\nu|\si\la)- P_{\la\si}^\ae (\mu\la|\si\nu),\\
 F_{\mu\nu}^\be= H_{\mu\nu}^{\rm core}  +
\sum_{\la} \sum_{\si} P_{\la\si}^T [(\mu\nu|\si\la)- P_{\la\si}^\be (\mu\la|\si\nu),
\label{fo}
\end{split}
\end{align}
and the $\ve_j^\az$, $\ve_j^\bz$, $\ve_j^\ae$, $\ve_j^\be$ being the energies for
the spin-isospin va-UHF orbitals. Because the $\vp_\mu$'s are eigenfunctions of $H^{\rm core}$, one has
$S_{\mu\nu}=\delta_{\mu\nu}$ for their overlaps. The two-body Coulomb matrix elements are given by
\begin{align}
\begin{split}
(\mu\nu|\si\la)=\frac{e^2}{\kappa} \int d\br_1 d\br_2 \vp^*_\mu(\br_1) \vp_\nu(\br_1) r^{-1}_{12}
\vp^*_\si(\br_2) \vp_\la(\br_2),
\end{split}
\end{align}
with $r^{-1}_{12} = 1/|\br_1-\br_2|$.

In Eq.\ (\ref{fo}), the partial density matrices are given by
\begin{align}
\begin{split}
 P_{\mu\nu}^\az=\sum_a^{N^\az} C_{\mu a}^\az (C_{\nu a}^\az)^*,\;
 P_{\mu\nu}^\bz=\sum_a^{N^\bz} C_{\mu a}^\bz (C_{\nu a}^\bz)^*,\\
 P_{\mu\nu}^\ae=\sum_a^{N^\ae} C_{\mu a}^\ae (C_{\nu a}^\ae)^*,\;
 P_{\mu\nu}^\be=\sum_a^{N^\be} C_{\mu a}^\be (C_{\nu a}^\be)^*,
\end{split}
\end{align}
the total-density matrix is defined as
 \begin{align}
 \begin{split}
 {\bf P}^T= {\bf P}^\az +  {\bf P}^\bz +  {\bf P}^\ae +  {\bf P}^\be,
 \end{split}
 \end{align}
and 
 \begin{align}
 \begin{split}
  N^\az +  N^\bz +  N^\ae +  N^\be = N.
 \end{split}
 \end{align}

The energy eigenvalues ($\ve_j^\az$, $\ve_j^\bz$, $\ve_j^\ae$, $\ve_j^\be$) and expansion coefficients 
($C_{\nu j}^\az$, $C_{\nu j}^\bz$, $C_{\nu j}^\ae$, $C_{\nu j}^\be$) are obtained via self-consistent 
solutions of Eqs.\ (\ref{popnes}).

We note that, unlike the valleytronic FCI \cite{yann22.3}, the va-UHF does not conserve the
total spin $\hat{\bf S}^2$ and total valley $\hat{\bf V}^2$ quantum numbers; it only conserves their 
projections $S_z$ and $V_z$. 

\section{{va-UHF} charge densities} 

\begin{figure}[t]
\centering\includegraphics[width=6.5cm]{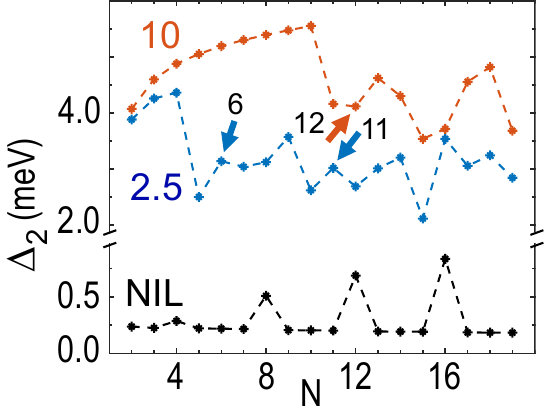}
\caption{
UHF second energy differences, $\Delta_2$. The curves associated with $\hbar \omega_y=2.5$ meV and
$\hbar \omega_y=10$ meV for $\kappa=11$ (strong interaction), as well as with the non-interacting 
limit (at maximum occupation per orbital, $\kappa=300$, and $\hbar \omega_y=2.5$ meV) are as noted. 
The arrows denote the transition sizes ($N^*$'s) between the multiple chains marked in Fig.\ 
\ref{uhfdens1} and Fig.\ \ref{uhfdens2}. 
}
\label{scddif}
\end{figure}

For $N=4-18$, fully spin and valley polarized ($S_z=V_z=N/2$), electrons 
and $\hbar \omega_y=2.5$ meV, Fig.\ \ref{uhfdens1} contrasts the UHF CDs at $\kappa=11$  (silicon, top 
rows) to those associated with the non-interacting limit (NIL) (at maximum occupation per orbital, 
$\kappa=300$, and $\hbar \omega_y=2.5$ meV, bottom rows). 
For all the strongly-interacting-electron cases, the UHF densities exhibit 
explicit configurations of $N$ well defined humps reflecting formation of pinned WMs, associated with
the regime of strong inter-electron correlations. In contrast, the NIL CDs conform to those expected 
from (Aufbau) shell closures in the $U(x,y)$ confinement, with a fourfold period (due to both 
the spin and valley) being clearly visible for $N=5-8$ [when the $(1_x,0_y)$ orbital of
the confining potential $U(x,y)$ is sequentially occupied (see also Fig.\ \ref{scddif}); here 
$(n_x,n_y)$ denotes the number of nodes in the $x$- and $y$-directions].

\textcolor{black}{
In Fig.\ \ref{s4} of Appendix \ref{axx}, we demonstrate that the CDs shown 
in Fig.\ \ref{uhfdens1} (see also Fig.\ \ref{s5}), calculated 
for the spin and valley fully polarized ($S_z=V_z=N/2$) electrons, are essentially identical to the CDs 
for the va-UHF states with minimal spin, $S_z$, and isospin, $V_z$, projections (0 for even $N$ or $\pm$1/2 
for odd $N$). This finding concurs with the magneto-spectroscopy measurements \cite{dzur23} where it has 
been found that the ``Jellybean'' quantum dot studied in these experiments lacks visible spin structure for
similar electron numbers as those investigated in the present paper.}
  
We recall that the strength of correlations is often expressed via the Wigner parameter 
\cite{yann99,yann07} $R_W={\cal E}_C/\delta$, defined as the ratio of a typical inter-electron 
repulsive Coulombic energy, ${\cal E}_C$, over a typical energy gap, $\delta$, in the single-particle 
spectrum of $U(x,y)$; naturally, the strong-interaction regime is expected for $R_W > 1$, accompanied
by WM formation. For the square-like 
confinement here, we take ${\cal E}_C$ as the two-body matrix element of the Coulomb interaction 
for two electrons occupying the nodeless lowest $(0_x,0_y)$ orbital, and for $\delta$ we take 
the difference between the energies of the $(0_x,0_y)$ and $(1_x,0_y)$ orbitals. For the EQD 
investigated here, we have $R_W=23.04$ for $\kappa=11$ and $\hbar \omega_y=2.5$ meV, 
$R_W=29.14$ for $\kappa=11$ and $\hbar \omega_y=10$ meV (strong interactions, case of the Si QD), and 
$R_W=0.84$ for $\kappa=300$ and $\hbar \omega_y=2.5$ meV (non-interacting limit).  

A prominent feature of the WM configurations is the successive formation with increasing $N$ of complex,
polymeric, multiple chain-like arrangements, reminiscent of the classical Wigner-crystalline chains 
associated with the equilibrium configurations of classical 
point charges \cite{piac04} in infinite-length wires. 
Naturally, the presence of the square-like edges here perturbs the perfect-chain formations of the 
infinite-length wires by forcing an accumulation of charges at the edges. The well-known zig-zag 
configuration, however, is visible away from the edges of the EQDs for $N=7-11$ (single zig-zag) and
for $N=14-18$ (double zig-zag).

We note that a single row appears for $N=1-5$, whilst a second row starts developing at $N=7$ with $N=6$ 
being a transitional stage. A third row starts developing at $N=11$, whilst transitional cases towards a 
fourth row appear at $N=17$ and $N=18$, with four electrons accumulating in a line along the 
$y$-direction at the edges. 

Fig.\ \ref{uhfdens2} displays the UHF CDs for fully spin and valley polarized $N=1-20$
electrons for a tighter transverse confinement with $\hbar \omega_y=10$ meV. These CDs exhibit again
organization with Wigner-chain-like features, the main difference from the $\hbar \omega_y=2.5$ meV case 
[see Fig.\ \ref{uhfdens1}] being a delay in the appearance (as a function of $N$) of the transition 
regions  between the multiple chains. In particular, the single chain transitions to a double zig-zag
chain at the region $N=12-14$, with the double chain starting to form from the edges 
\footnote{This behavior contrasts with that in harmonic confinements along the lateral $x$-direction, 
where the zig-zag chain starts forming at the center of the linear chain, a fact well known from 
classical calculations in the literature of trapped heavy ions; see, e.g., Refs.\ 
\cite{schi93,ejte15,yan16}}, a behavior related to the accumulation of charge at the sharp square-like 
edges of the confinement along the $x$-direction. Note that, due to this delay, the third row does not 
appear in Fig.\ \ref{uhfdens2} in the range $N=1-20$; for $\hbar \omega_y=10$ meV, a third row is 
expected to develop at larger values of $N$.

\begin{figure}[t]
\centering\includegraphics[width=7.5cm]{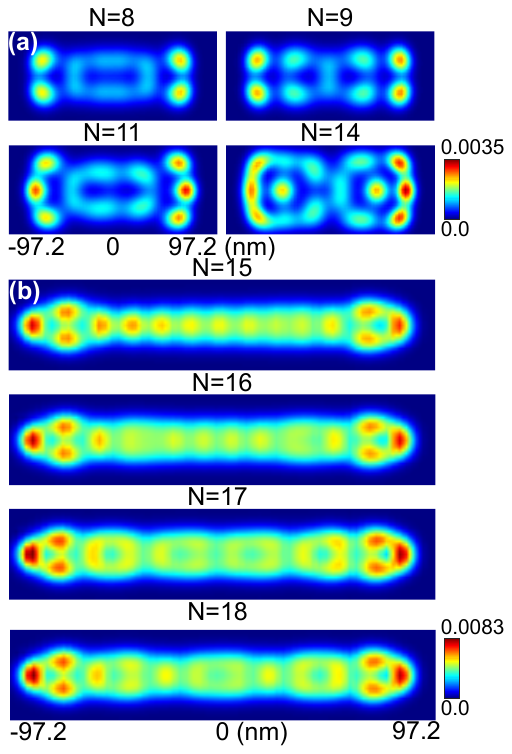}
\caption{
Parity-restored CDs for a silicon EQD. 
(a) transverse confinement of $\hbar \omega_y=2.5$ meV. $N=8,9,11,14$ fully spin and valley 
polarized electrons.  
(b) transverse confinement of $\hbar \omega_y=10$ meV. $N=15-18$ fully spin and valley polarized
electrons. 
In both cases, $\kappa=11$ appropriate for silicon (strong interaction). The symmetry-broken UHF 
CDs, corresponding to the restored ones shown in (a) and (b), are given in 
Figs.\ \ref{uhfdens1} and \ref{uhfdens2}, respectively. 
Obviously, only symmetry-broken UHF wave-functions undergo symmetry restoration. 
CDs in units of 1/nm$^2$.
}
\label{parres}
\end{figure}

\section{Restoration of parity along the transverse $y$-direction}
The broken-symmetry UHF solutions violate a fundamental axiom of quantum mechanics, namely, that the
single-particle CDs must preserve the symmetries of the many-body Hamiltonian. Here we take a first 
step in rectifying this UHF inadequacy by restoring the parity symmetry along the transverse 
$y$-direction, which is associated with a symmetric harmonic confinement $U_T(-y)=U_T(y)$; note that 
the confinement along $x$ was taken to be slightly anisotropic.

To restore the $y$-parity symmetry, we apply on the UHF Slater determinant (SD), $\uhf$, the projection 
operator
\cite{yann07,yann21,paca70}
\begin{align}
\hpi=\frac{1}{2} (1+p \cpy),
\label{prt}
\end{align}
with $p=\pm 1$. $\cpy$ is the many-body parity operator, which inverts about the $x$-axis the 
$y$-coordinates for all electrons, namely, $\cpy=\prod_{i=1}^N \hat{P}^{(i)}_y$.

The charge density associated with the projected wave function $\puhf=\hpi \uhf$ is given by
\begin{align}
\rho_{\rm PUHF}(\br) = {\cal N} \langle \uhf | \hpi \sum_{i=1}^N \delta (\br_i-\br) \hpi 
| \uhf \rangle,
\label{puhfdens}
\end{align}
where the factor ${\cal N}$ imposes normalization to $N$. 
The expansion of the symmetry-restored charge density in Eq.\ (\ref{puhfdens}) is explicitly displayed 
in Eq.\ (\ref{oper}) of Appendix \ref{ayy}.

The projected total energy is given by \cite{yann07,yann21}
\begin{align}
E_{\rm PUHF}= \frac{\langle \uhf | H \hpi | \uhf \rangle} {\langle \uhf | \hpi | \uhf \rangle}.
\label{puhfener}
\end{align}  
Eqs.\ (\ref{puhfdens}) and (\ref{puhfener}) are calculated using the property $\hat{P}^{(i)}_y Y_ny)=
\varpi_n Y_n(y)$, where $\varpi_n=(-1)^n$, $n=0,1,2,\ldots$ and $Y_n(y)$ are the good-parity 
eigenfunctions  of $U_T(y)$.

Several CD results of this parity restoration are displayed in Fig.\ \ref{parres}; 
it is apparent that the  zig-zag  motif gets obliterated by the parity restoration.
Whether a zig-zag or a symmetry-restored CD will be the actual finding in a quasi-1D system 
may depend in practice on several additional factors, including the 
presence of impurities, the length of the system, as well as the value of the effective mass. Indeed, 
the longer the system, the higher the expectation for a broken symmetry CD. Interestingly, because of 
the much larger mass compared to electrons, the CDs of trapped heavy ions are routinely found 
experimentally to exhibit broken-symmetry structures, including zig-zag chains. Nevertheless, in spite 
of the large mass of the ions, the superposition (or entanglement) of both the ``zig'' and ``zag'' 
configurations is quantum mechanically allowable, and recently it has been observed experimentally 
\cite{ejte23}.  

\begin{figure}[t]
\centering\includegraphics[width=7.2cm]{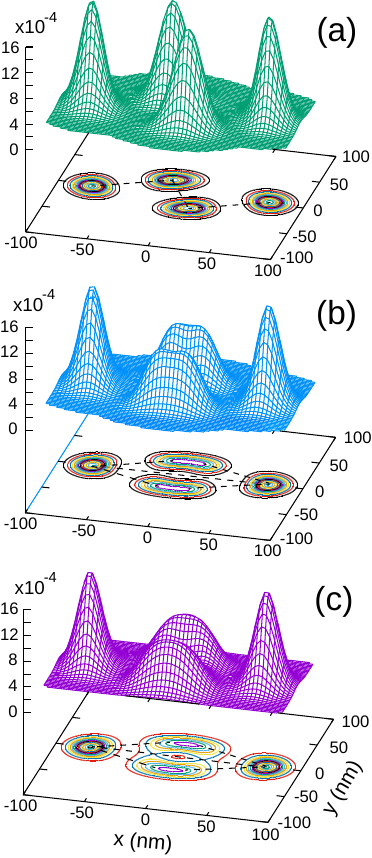}
\caption{
Comparison of (a) symmetry-broken UHF, (b) $y$-parity-restored UHF, and (c) FCI CDs for the associated
ground states of $N=4$ electrons at zero magnetic field in a single elliptic quantum dot, with an effective
mass $m^*=0.067m_e$ and a potential confinement specified by $\hbar \omega_x=3.00$ meV and
$\hbar \omega_y=5.64$ meV.
The dielectric constant used was $\kappa=1$. In (a) and (b), $S_z=0$. In (c) $S=S_z=0$.
Lenghts in units of nm. Charge densities in units of 1/nm$^2$. The charge densities are normalized to
the total number of fermions $N=4$. The dashed lines are a guide to the eye; in (b) and (c), they
delineate two mirror trapezoids.}
\label{s6}
\end{figure}


Even though it is expected that (due to the much smaller electronic mass) the actual CDs of electrons
in EQDs and finite length wires would conform to those produced through the symmetry-restoration 
correction, a promising experimental opportunity emerged recently in the case of 1D domain walls in 
TMD bilayers  \cite{crom24}, where rapid progress in STM imaging could differentiate between 
symmetry-broken and symmetry-preserving CDs.

\section{Second energy differences}

The second difference of the total energies, $\Delta_2=E(N+1)+E(N-1)-2E(N)$, corresponding to the va-UHF 
electronic configurations shown in Figs.\ \ref{uhfdens1} and \ref{uhfdens2} are shown in Fig.\ \ref{scddif}.
Also shown is the non-interacting case (at maximum occupation per orbital, $\kappa=300$, and 
$\hbar \omega_y=2.5$ meV), which exhibits shell closures at $N = 4$, 8, 12, and 16 according to 
the Aufbau principle, with the period of 4 resulting from both the spin and valley degrees of freedom. 
This non-interacting behavior contrasts sharply with that of the strong-interaction cases, where polymeric  
WM chains are formed. We note that $\Delta_2$ is proportional to the capacitance of the EQD.

\section{AN EXAMPLE OF A COMPARISON WITH full configuration interaction}

In Fig.\ \ref{s6}, we present a case of a comparison between UHF and parity-restored UHF charge
densities with those from the corresponding FCI \cite{yann22.2,yann22.3} calculation.
We choose the case of an elliptic external 2D potential that binds $N=4$ conduction electrons for a 
one-band (no valley present) semiconductor material.

The UHF CD in Fig.\ \ref{s6}(a) displays a broken-symmetry zig-zag configuration. However,
restoration of the parity along the $y$-axis in Fig.\ \ref{s6}(b) (which also restores the $x$-parity in
this example) produces a configuration of two mirror trapezoids, which is formed by the spreading out
(along the $x$-axis) of the two middle sharp single-electron humps in panel Fig.\ \ref{s6}(a). In this
context, note the reduction in height of the corresponding broad humps in Fig.\ \ref{s6}(b). The
parity-restored UHF CD in Fig.\ \ref{s6}(b) is in good agreement with the FCI CD in Fig.\ \ref{s6}(c).
Naturally, the FCI solution displays a certain degree of additional relaxation effects in the CD humps.

\section{Discussion and Conclusions} 

We introduced a generalization (referred to as va-UHF) of the 
unrestricted Hartree-Fock methodology \cite{szabo,yann99,yann02.2,yann07}
that incorporates the valley degree of freedom on an equal footing with the electronic spin. Furthermore, 
we explored an augmenting symmetry (here, parity)-restoration step applied to the symmetry-broken va-UHF 
(going beyond the single determinant description) that amplifies and improves the quantum mechanical 
description of the considered nanosystem \cite{yann02.2,yann07,yann21}. 
This new va-UHF methodology is able to effectively and economically investigate the 
physics of double-quantum-dot-based Si qubits for a number of charge carriers much larger 
than the number accessible with FCI approaches  
\cite{lowd55,shav98,yann03,yann07,yann22.2,yann22.3}. 
\textcolor{black}{
The close agreement between the results of charge densities obtained via the symmetry-restored (SR)-UHF 
methodology and those evaluated by full configuration (exact diagonalization) computations is illustrated 
in Fig.\ \ref{s6}).}

\begin{figure}[t]
\centering\includegraphics[width=7.5cm]{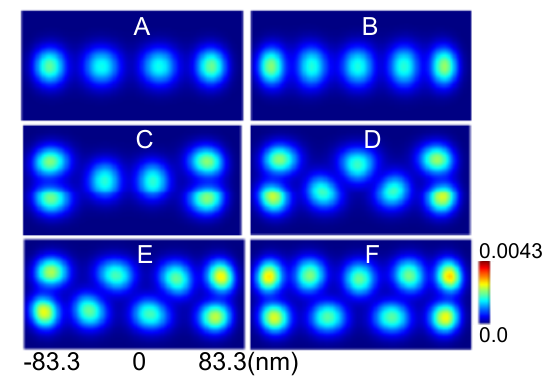}
\caption{
Ground-state va-UHF charge densities for $N=4-9$ fermions in the same elongated-dot confinement as
employed in Fig.\ \ref{uhfdens1} of the main text, associated with states with minimum spin and isospin 
polarizations,
as follows: (A) $S_z = 0$, $V_z = 0$, $N = 4$. (B) $S_z =1/2$, $V_z =1/2$, $N = 5$.
(C) $S_z = 0$, $V_z = 0$, $N = 6$. (D) $S_z =1/2$, $V_z = -1/2$, $N = 7$. (E) $S_z = 0$, $V_z = 0$, $N = 8$.
(F) $S_z =1/2$, $V_z =1/2$, $N = 9$.
The square-like potential $U_L(x)$ along the $x$-direction is given by Eq.\ (\ref{potx}) in the main 
text with $x_0=50$ nm and $\xi=0.1$ meV. The harmonic confinement $U_T(x)$ along the $y$-direction has
$\hbar \omega_y=2.5$ meV. Other parameters used: effective mass $m^*=0.19m_e$ and dielectric constant $\kappa=11$. Charge densities in units of 1/nm$^2$.
}
\label{s4}
\end{figure}

\begin{figure}[t]
\centering\includegraphics[width=7.5cm]{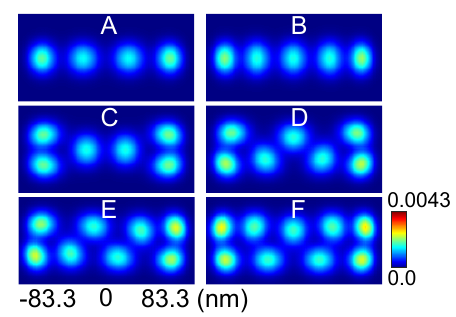}
\caption{
The fully spin and isospin polarized, $(S_z=V_z=N/2)$, va-UHF CDs for $N=4-9$ to be contrasted with those in
Fig.\ \ref{s4}. Parameters are the same as in the corresponding frames in Fig.\ \ref{s4}.
CDs in units of 1/nm$^2$.}
\label{s5}
\end{figure}

A pioneering va-UHF application was presented here for the case of an elongated \cite{kuem23} 
(referred to also as a jellybean \cite{dzur23}) Si QD. EQDs have been proposed as long-distance 
couplers between solid-state qubits, enabling thus an effective solution to the problem of 
scalability \cite{kuem17,kuem18,dzur23,kuem23}.
Our calculations for a typical Si EQD revealed a strongly correlated regime as a particular case 
of Wigner molecularization  \cite{yann99,yann02.2,yann07,yann11}, 
with the electrons organized in polymeric multi-chain arrangements 
which are finite-size analogs of the infinite Wigner chains considered earlier
\cite{piac04,matv08,pepp18,pepp21}. In this context, we note that our results agree with the
experimental findings of Ref.\ \cite{kuem23} that the charge is "well distributed" along the
long lateral axis of the EQD, whereas Ref.\ \cite{dzur23} concluded that the electrons
bunch together forming several separated QDs. Instead of separated QDs, our investigations suggest that
the fragmentation in the transverse direction associated with the formation of multiple chains may 
explain the variety of conductance behaviors reported in Ref.\ \cite{dzur23}. 
\textcolor{black}{
Conductance calculations between quantum dots coupled to the elongated coupler element, based on  
the many-body restored-parity va-UHF wave functions, will constitute a promising follow up step of 
the present paper.}  

Finally, we note that the va-UHF findings could be further confirmed with a newly developing 
experimental platform associated with 1D domain walls in twisted TMD bilayers \cite{crom24}.

\section{acknowledgments}

This work has been supported by a grant from the Air Force Office of Scientific Research (AFOSR)
under Grant No. FA9550-21-1-0198. Calculations were carried out at the GATECH Center for
Computational Materials Science.

\appendix

\begin{figure}[t]
\centering\includegraphics[width=8.0cm]{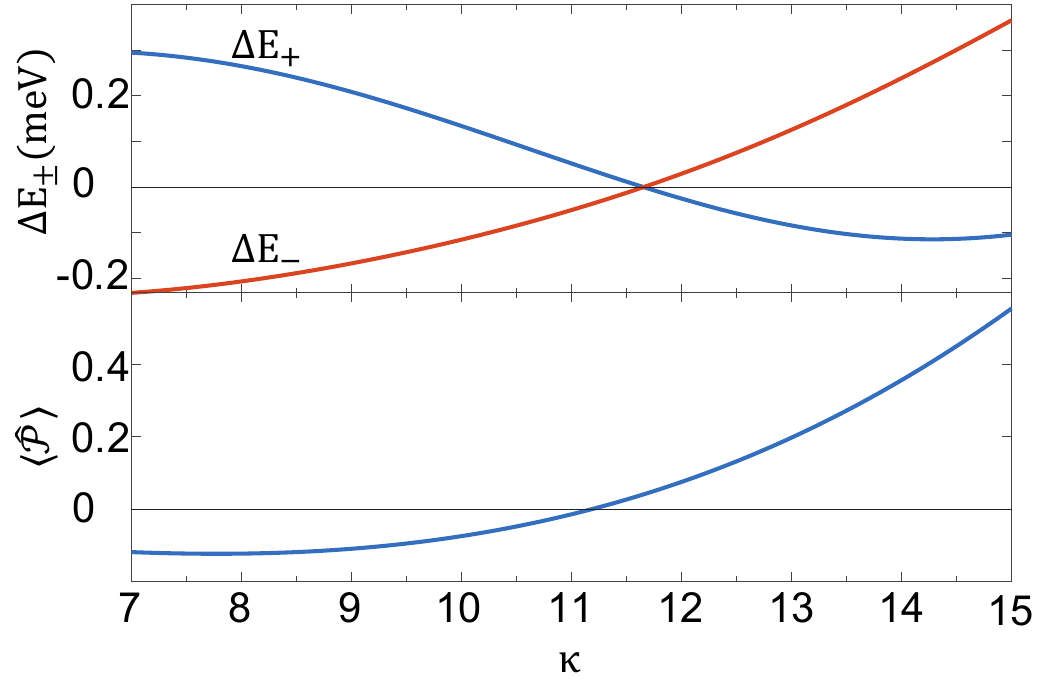}
\caption{
The energy difference (top), $\Delta E_\pm = E_{\rm PUHF} - E_{\rm UHF}$ and expectation value of the
parity operator (bottom), $\langle {\rm UHF} | {\cal P} | {\rm UHF} \rangle$, for $N=7$ fully spin and valley
polarized electrons as a function of the dielectric conctant $\kappa$. The $\pm$ subscript corresponds to
the two eigenvalues $p=\pm 1$ entering in the definition of the parity projection operator; 
see Eq.\ (\ref{prt}) of the main text. The parameters used are: $m^*=0.19 m_e$, $\hbar \omega_y=2.5$ meV.
}
\label{s3}
\end{figure}

\section{SOME CASES OF CHARGE DENSITIES FOR va-UHF STATES WITH MINIMUM SPIN AND ISOSPIN POLARIZATION}
\label{axx}

In Fig.\ \ref{s4}, we demonstrate that the va-UHF CDs shown in Fig.\ \ref{uhfdens1}, calculated for the
spin and valley fully polarized electrons, are essentially identical to the CDs for the va-UHF states with
minimal spin, $S_z$, and isospin, $V_z$, projections; see the caption for the specific values. This finding
concurs with the magneto-spectroscopy measurements \cite{dzur23} where it has been found that the
``Jellybean'' quantum dot studied in these experiments lacks visible spin structure for similar electron
numbers as those investigated in the present paper.
In order to facilitate the direct comparison, we display in Fig.\ \ref{s5} the corresponding
fully spin and isospin polarized ($S_z=V_z=N/2$) va-UHF CDs.

\section{AN EXAMPLE OF CALCULATIONS OF PROJECTED ENERGIES}

According to Eqs.\ (\ref{prt})-(\ref{puhfener}) of the main text, there are always two projected states, 
$\Psi_{\rm PUHF}$, with good parity eigenvalues $p=\pm 1$ for each mixed-parity state, $\Psi_{\rm UHF}$.
The top panel of Fig.\ \ref{s3} demonstrates that one of the symmetry-restored states has always an
energy lower (at most equal to) than the broken symmetry UHF state. The bottom panel of Fig.\ \ref{s3}
displays the expectation value of the parity operator, $\langle {\rm UHF} | {\cal P} | {\rm UHF} \rangle$, 
which varies between -1 and +1. Note that, as a function of a increasing  dielectric constant $\kappa$, the
negative-parity projected ground state is replaced by the positive-parity one, following closely a
similar behavior by the expectation value.

\section{CLARIFICATION CONCERNING THE COMPACT EXPRESSION IN EQ. (13) OF THE MAIN TEXT}
\label{ayy}

Denoting the va-UHF Slater determinant as $\Psi(x,y)$ (we drop the subscript UHF here), its mirror image
about the $x$ axis is given by $\Psi(x,-y)$, and the $y$-parity restored wave function is
$\propto \Psi(x,y)+p \Psi(x,-y)$, with $p=\pm1$. Then, because the Slater determinant
$\Psi(x,-y)$ is in general not orthogonal to $\Psi(x,y)$, the
expectation value of an operator ${\cal O}$ is given by
\begin{widetext}
\begin{align}
\frac{\langle \Psi(x,y)| {\cal O}| \Psi(x,y) \rangle +
p \langle \Psi(x,y)| {\cal O}| \Psi(x,-y) \rangle +
p \langle \Psi(x,-y)| {\cal O}| \Psi(x,y) \rangle +
 \langle \Psi(x,-y)| {\cal O}| \Psi(x,-y) \rangle}
{\langle \Psi(x,y)| \Psi(x,y) \rangle + p \langle \Psi(x,y)| \Psi(x,-y) \rangle +
p \langle \Psi(x,-y)| \Psi(x,y) \rangle + \langle \Psi(x,-y)| \Psi(x,-y) \rangle}.
\label{oper}
\end{align}
\end{widetext}
The operator associated with the charge density is a one-body operator,
$\sum_{i=1}^N \delta(\br - \br_i)$.

\nocite{*}
\bibliographystyle{apsrev4-2}
\bibliography{mycontrols,elongated_QD_PRApplied}

\end{document}